**Ultrasonic Attenuation and Speed of Sound of Cornstarch Suspensions**

Benjamin L. Johnson, Mark R. Holland, James G. Miller, and Jonathan I. Katz

Department of Physics

Washington University in St. Louis

St. Louis, MO, 63130

Running Title: Ultrasonic Measurements of Cornstarch Suspensions

**Abstract**

The goal of this study is to contribute to the physics underlying the material properties of suspensions that exhibit shear thickening through the ultrasonic characterization of suspensions of cornstarch in a density-matched solution. Ultrasonic measurements at frequencies in the range of 4 MHz to 8 MHz of the speed of sound and the frequency-dependent attenuation properties are reported for concentrations of cornstarch in a density-matched aqueous (cesium chloride brine) suspension, ranging up to 40% cornstarch. The speed of sound is found to range from 1483 m/s ± 10 m/s in pure brine to 1765 m/s ± 9 m/s in the 40% cornstarch suspension. The bulk modulus of a granule of cornstarch is inferred to be $1.2(\pm 0.1)*10^{10}$ Pa. The attenuation coefficient at 5 MHz increases from essentially zero in brine to 12.0 dB/cm ± 1.2 dB/cm at 40% cornstarch.

## I. INTRODUCTION

Aqueous suspensions of cornstarch are known for displaying discontinuous shear thickening. The mechanisms underlying this behavior are an active subject of research.[1-7] Suspensions that shear thin or thicken are of fundamental interest and are also of potential practical importance. Flowing blood, which consists of liquid plasma and (soft-solid) formed elements, can exhibit shear thinning behavior.[8] Shear thickening fluids have been considered as a means of capping blown-out oil wells[9] and as a component of body armor.[10] As indicated in the Discussion section below, the ultrasonic characterization of static cornstarch suspensions provides a useful first step toward using ultrasound to study the physics of these suspensions under shear.

Ultrasound has been used to characterize cornstarch (and starches in general) for a number of different applications. The food industry has used ultrasound to examine starches and flours for at least 25 years in order to study gelatinization.[11-20] Ultrasound has also been used to characterize dense slurries similar to the cornstarch suspensions.[21-23] Several papers have examined the flow properties of suspensions of cornstarch using ultrasonic methods.[24-26] Cornstarch has also been used as a scatterer in ultrasonic phantoms.[27-28] Koltsova et al. characterized cornstarch in relatively low concentrations of non-density-matched suspensions.[29-30]

The present investigation reports ultrasonic (4 to 8 MHz) measurements of the speed of sound and the frequency-dependent attenuation for concentrations of cornstarch ranging from 0% to 40% in a density-matched aqueous suspension.

## II. METHODS

### A. Sample preparation

Cornstarch granules (Sigma Aldrich, St. Louis, MO) have a density of approximately 1.6 g/cm$^3$ and settle too rapidly in pure water to permit quantitative experimentation. Settling is slowed by the use of a density-matched fluid.[5,7] It was found 51.5% (mass fraction) cesium chloride brine in water to yield the best match. This is close, but not identical, to the concentration used in other studies, perhaps reflecting variability of the properties of cornstarch, a natural product that absorbs and desorbs environmental water. The density of the 51.5% brine was measured to be 1.62 g/mL ± 0.01 g/mL using a precision pipette (Biohit Corporation, Laippatie, Helsinki) and a XS105 Dual Range Balance (Mettler Toledo, Greifensee, Switzerland), consistent with standard results.[31] The density of the cornstarch granules falls within the range of values measured in the literature.[7,32]

Cornstarch suspensions were measured in a cylindrical plastic (Lexan™) holder of diameter approximately 45 mm fitted with thin plastic wrap (Saran Wrap™) windows on each end. The ultrasonic path length through the suspension was approximately 12 mm.

### B. Speed of sound measurement

The speed of sound in the cornstarch suspensions was measured in reflection mode using a time-of-flight technique using the signals reflected from the front and back surfaces of the sample and from a steel plate reflector with and without the sample present.[33,34] The experimental setup is shown in Figure 1. The sample holder containing the starch was immersed in a host medium of water. The data were acquired using a focused piezoelectric transducer (Panametrics V309 transducer with a nominal center frequency of 5 MHz, a 0.5 inch (12.7 mm) diameter and a 2 inch (50.8 mm) point target focus; Panametrics, Waltham, MA) driven by a Panametrics 5800 Pulser/Receiver. The fractional bandwidth of the transducer was 80%. The signal was received by the same transducer, amplified by the Pulser/Receiver, and digitized at 8 bits by the TDS5052 Digital Oscilloscope (Tektronix, Beaverton, OR). The received signals were averaged over 256 pulses.

The speed of sound in the sample was determined using

$$v_s = v_h \left[ 1 + \frac{t_{ref} - t_{samp}}{t_{BW} - t_{FW}} \right] \qquad (1).$$

where $v_h$ is the velocity in the host medium, $t_{ref}$ is the roundtrip time for the reference measurement without the sample, $t_{samp}$ is the roundtrip time with the sample interposed, $t_{FW}$ is the roundtrip time for the signal to reflect off of the front wall of the sample and return to the transducer, and $t_{BW}$ is the roundtrip time for the signal to reflect off of the back wall of the sample and return to the transducer. The pertinent times were determined from the maximum of the magnitude of the analytic signal. The magnitude of the analytic signal was obtained as

$$|x_a(t)| = \sqrt{[x(t)]^2 + [H\{x(t)\}]^2} \qquad (2)$$

where $x_a(t)$ is the analytic signal of some function of time $x(t)$ and $H\{x(t)\}$ is the Hilbert transform of $x(t)$.[35] An advantage of this method is that it does not require an independent measurement of the sample thickness.

The sample thickness was determined from the same timing measurements, and is

$$\ell = \frac{v_h}{2}\left[\left(t_{ref} - t_{samp}\right) + \left(t_{BW} - t_{FW}\right)\right] \qquad (3).$$

The host velocity is calculated from a known polynomial relationship between the temperature of water and the speed of sound in water.[36] The temperature was monitored throughout the experiments and did not vary more than 0.2 °C. This is approximately equivalent to variation of 0.5 m/s in the speed of sound in the water.

    The speed of sound was determined for eight samples prepared and measured at each concentration (10%, 20%, 30%, and 40%) of cornstarch in a density-matched 51.5% cesium chloride solution. In order to characterize the medium in the absence of shear thickening, the concentrations studied in this manuscript range from well below the point at which shear thickening has been observed to the range (40%) beyond which shear thickening has been reported.[5] The speed of sound was also measured for two 51.5% cesium chloride solutions with no cornstarch added (0% cornstarch).

**C. Attenuation measurements**

The frequency dependent attenuation coefficient was determined using the log-spectral subtraction method.[37] The experimental setup is shown in Figure 1. The reference measurement was made by recording the reflection of a signal from a steel reflector placed at the focus of the transducer. The sample measurement was also made with the focus of the transducer on the steel reflector but with the sample placed in between the transducer and the reflector. The attenuation coefficient was measured over a bandwidth of 4 MHz to 8 MHz for eight samples at each concentration.

The frequency dependent attenuation coefficient in units of dB/cm was determined with the log spectral subtraction technique,

$$\alpha_s^{dB}(f) = \frac{1}{2\ell}\left[10\log_{10}\left(|\tilde{V}_r|^2\right) - 10\log_{10}\left(|\tilde{V}_s|^2\right) + 10\log_{10}\left\{\left(T_{w\to s\to c}^{I}(f)\right)^2 \left(T_{c\to s\to w}^{I}(f)\right)^2\right\}\right] \quad (4).$$

The reference and sample power spectra, $|V_r|^2$ and $|V_s|^2$, are determined from the measurements of the reference and sample signals. The log spectral subtraction of the two power spectra detailed in the first term on the right hand side of Equation 4 results in the signal loss. The signal loss is compensated by the intensity transmission coefficients, $T_{w\to s\to c}^{I}$, and then divided by the sample thickness, $\ell$, in order to arrive at the attenuation coefficient. The effect of diffraction on the measurement of the attenuation coefficient as a function of frequency was investigated experimentally using an approach suggested by Wu and Kaufman.[38] For all suspensions studied, the diffraction-correction was negligible.

**III. RESULTS**

The speed of sound in the cornstarch suspensions in 51.5% cesium chloride solutions is plotted as a function of cornstarch concentration in 21 and summarized in Table 1. The speed of sound in the cornstarch suspensions increased progressively from 0% to 40% cornstarch concentration.

The measurements of the attenuation coefficient as a function of frequency are shown in Figure 3. The figure displays the individual measurements of the attenuation coefficient for each of the eight independent samples at each concentration. The scatter among nominally identical samples is an indication of the reproducibility of the measurements. Potential sources of this scatter include sample heterogeneity resulting from sedimentation (density-matching is not perfect and starch granules are unlikely to be of identical density), imperfect initial mixing, and acoustic refraction as a result of heterogeneities. The attenuation coefficient increased with increasing concentration of cornstarch. Table 1 displays the average attenuation coefficients at selected frequencies for each of the concentrations of cornstarch.

## IV. DISCUSSION

This study ultrasonically characterized suspensions of cornstarch in a density-matched aqueous solution at concentrations up to 40%. Both the speed of sound and the frequency dependent attenuation coefficient increase progressively with cornstarch concentration.

The measurements of the speed of sound were used to estimate the bulk modulus in the cornstarch granule itself. This method has earlier been used to determine the compressibility of oil droplets[39] and the compressibility of red blood cells[40] from measurements of suspensions. For a suspension of two components that do not react chemically with each other, the speed of sound is related to the bulk moduli and mass densities of the components by

$$c = \sqrt{\frac{K_1 K_2}{\left[f_1 K_2 + (1-f_1)K_1\right]\left[f_1 \rho_1 + (1-f_1)\rho_2\right]}} \qquad (5)$$

where $\rho_1$ is the density of the suspending medium (51.5% cesium chloride solution), $\rho_2$ is the density of the granule of cornstarch, $K_1$ is the bulk modulus of the suspending medium, $K_2$ is the bulk modulus of the granule of cornstarch, c is the speed of sound in the suspension, and $f_1$ is the volume fraction of the suspending medium.[41] In this work, $\rho_1 = \rho_2$ because of the use of a density matched solution. The speed of sound and bulk modulus of a material are related by

$$c = \sqrt{\frac{K}{\rho}} \qquad (6).$$

Because the starch granules are very small compared to the ultrasonic wavelength, their response to shear may be neglected. Equation 5 can be re-expressed to determine the speed of sound in the cornstarch granules, $c_2$, where $c_1$ is the speed of sound in the host medium

$$c_2 = \sqrt{\frac{c^2 c_1^2 (1 - f_1)}{c_1^2 - c^2 f_1}} \qquad (7).$$

Table 2 shows the calculated speed of sound in the cornstarch granules at suspension concentrations of 20%, 30%, and 40%. Results for the 10% concentration are not shown because the denominator of Equation 7 vanishes at low concentrations, making the inferred $c_2$ sensitive to small experimental errors. Values of the bulk modulus in the cornstarch granules were inferred from the calculated speed of sound at each concentration using Equation 6 and a density of 1.62 g/cm³.

The bulk modulus of cornstarch shown in Table 2 is consistent (to within the expected accuracy of measurement) for the 20%, 30%, 40% concentrations. The bulk moduli for an extruded wheat starch gel containing 34% water and for wheat flour dough containing about 50% water were both reported to be approximately $5*10^9$ Pa.[17,18] The value for the bulk modulus reported is for a pure cornstarch granule, whereas the estimates from the literature were for a gelatinized starch and a flour dough both containing significant amounts of water, so the difference is not unexpected.

The data reported in Table 1 and Figure 3 indicate that the attenuation varies slightly more steeply than the first power of the frequency. This lies between the low frequency ($f^2$) and high frequency ($f^{0.5}$) exponents of classical theory.[42] For cornstarch granules with typical radii of 7 microns, insonification at 4-8 MHz is in the low frequency regime; the measured attenuation is about an order of magnitude greater than expected on the basis of viscous and thermal dissipation. It is plausible that the dominant source of dissipation is internal relaxation in the cornstarch. Internal relaxation typically gives an attenuation proportional to frequency, a somewhat, but not enormously, weaker dependence than what was measured.

The attenuation was found to vary slightly more steeply than the first power of the starch concentration. Theory predicts linear proportionality for viscous and thermal dissipation in the dilute limit, but this is not expected to be quantitatively applicable at the large concentrations of our experiments. Linear proportionality would also be expected for internal relaxation in the starch granules. At higher concentrations the granules interact acoustically and their behavior is expected to be complex.

## V. CONCLUSIONS

The goal of this study was to measure the material properties of suspensions of cornstarch in a density-matched solution. Ultrasonic measurements of the speed of sound and the frequency-dependent attenuation properties are reported for concentrations of cornstarch in a density-matched aqueous suspension ranging from 0% cornstarch to 40% cornstarch. The speed of sound and the frequency dependent attenuation coefficient were both found to increase progressively with cornstarch concentration. Results of these measurements were employed to determine the bulk modulus of the anhydrous cornstarch granule, a quantity that is not readily measured in any other way.

**Acknowledgements**

This study was supported in part by American Chemical Society/Petroleum Research Fund (ACS/PRF) #51987-ND9.

**References**

[1] E. Brown and H.M. Jaeger. "Dynamic jamming point for shear thickening suspensions." Phys. Rev. Lett. **103**, 086001 (2009).

[2] E. Brown, N.A. Forman, C.S. Orellana, H. Zhang, B.W. Maynor, D.E. Betts, J.M. DeSimone, and H.M. Jaeger. "Generality of shear thickening in dense suspensions." Nat. Mat. **9**, 220-224 (2010).

[3] E. Brown, H. Zhang, N.A. Forman, B.W. Maynor, D.E. Betts, J.M. DeSimone, and H.M. Jaeger. "Shear thickening and jamming in densely packed suspensions of different particle shapes." Phys. Rev. E **84**, 031408 (2011).

[4] X. Cheng, J.H. McCoy, J.N. Israelachvili, and I. Cohen. "Imaging the microscopic structure of shear thinning and thickening colloidal suspensions." Science **333**, 1276-1279 (2011).

[5] A. Fall, N. Huang, F. Bertrand, G. Ovarlez, and D. Bonn. "Shear thickening of cornstarch suspensions as a reentrant jamming transition." Phys. Rev. Lett. **100**, 018301 (2008).

[6] A. Fall, A. Lemaître, F. Bertrand, D. Bonn, and G. Ovarlez. "Shear thickening and migration in granular suspensions." Phys. Rev. Lett. **105**, 268303 (2010).

[7] F. Merkt, R. Deegan, D. Goldman, E. Rericha, and H. Swinney. "Persistent holes in a fluid." Phys. Rev. Lett. **92**, 184501 (2004).

[8] S. Chien, S. Usami, R.J. Dellenback, and M.I. Gregersen. "Shear-dependent deformation of erythrocytes in rheology of human blood." Am. J. Physiol. **219**, 136-142 (1970).

[9] P. Beiersdorfer, D. Layne, E. Magee, and J.I. Katz. "Viscoelastic suppression of gravity-

driven counterflow instability." Phys. Rev. Lett. **106**, 058301 (2011).

[10]Y. Lee, E. Wetzel, and N. Wagner. "The ballistic impact characteristics of kevlar woven fabrics impregnated with a colloidal shear thickening fluid." J. Mater. Sci. **38**, 2825-2833 (2003).

[11]J.M. Alava, S.S. Sahi, J. García-Alvarez, A. Turó, J.A. Chávez, M.J. García, and J. Salazar. "Use of ultrasound for the determination of flour quality." Ultrason. **46**, 270-276 (2007).

[12]C. Aparicio, P. Resa, L. Elvira, A.D. Molina-Garcia, M. Martino, and P.D. Sanz. "Assessment of starch gelatinization by ultrasonic and calorimetric techniques." J. Food Eng. **94**, 295-299 (2009).

[13]L.A. Cobus, K.A. Ross, M.G. Scanlon, and J.H. Page. "Comparison of ultrasonic velocities in dispersive and nondispersive food materials." J. Agric. Food Chem. **55**, 8889-8895 (2007).

[14]J. García-Álvarez, J. Salazar, and C.M. Rosell. "Ultrasonic study of wheat flour properties." Ultrason. **51**, 223-228 (2011).

[15]J. Garcia-Alvarez, J.M. Alava, J.A. Chavez, A. Turo, M.J. Garcia, and J. Salazar. "Ultrasonic characterisation of flour-water systems: A new approach to investigate dough properties." Ultrason. **44**, Supplement, e1051 - e1055 (2006).

[16]L. Lehmann, E. Kudryashov, and V. Buckin. "Ultrasonic monitoring of the gelatinisation of starch." Prog. in Coll. and Polym. Sci. **123**, 136-140 (2004).

[17]C. Létang, M. Piau, C. Verdier, and L. Lefebvre. "Characterization of wheat-flour-water doughs: a new method using ultrasound." Ultrason. **39**, 133-141 (2001).

[18]F. Lionetto, A. Maffezzoli, M.A. Ottenhof, I.A. Farhat, and J.R. Mitchell. "Ultrasonic investigation of wheat starch retrogradation." J. Food Eng. **75**, 258 - 266 (2006).

[19]M. Povey, and A. Rosenthal. "Technical note: ultrasonic detection of the degradation of starch by alpha-amylase." J. Food Tech. **19**, 115-119 (1984).

[20]J. Salazar, J.M. Alava, S.S. Sahi, A. Turo, J.A. Chavez, and M.J. Garcia. "Ultrasound measurements for determining rheological properties of flour-water systems."  Proc. IEEE Ultrason. Ferroelectr. Freq. Control, 877-880 (2002).

[21]V. Stolojanu, and A. Prakash. "Characterization of slurry systems by ultrasonic techniques." Chem. Eng. Jour. **84**, 215-222 (2001).

[22]C. Sung, Y. Huang, J. Lai, and G. Hwang. "Ultrasonic measurement of suspended sediment concentrations: an experimental validation of the approach using kaolin suspensions and reservoir sediments under variable thermal conditions." Hydro. Proc. **22**, 3149-3154 (2008).

[23]M. Xue, M. Su, L. Dong, Z. Shang, and X. Cai. "An investigation on characterizing dense coal water slurry with ultrasound: theoretical and experimental methods." Chem. Eng. Comm. **197**, 169-179 (2009).

[24]B. Birkhofer, S. Jeelani, B. Ouriev, and E. Windhab. "In-line characterization and rheometry of concentrated suspensions using ultrasound." *Proc. Ultrason. Doppler Method for Fluid Mech. and Fluid Eng. Symp.*, 65-68 (2004).

[25]B. Ouriev, and E. Windhab. "Rheological study of concentrated suspensions in pressure driven shear flow using a novel inline ultrasound Doppler method." J. Experim. in Fluids **32**, 204-211 (2002).

[26] B. Ouriev, and E. Windhab. "Novel ultrasound based time averaged flow mapping method for die entry visualization in flow of highly concentrated shear-thinning and shear-thickening suspensions." Measure. Sci. and Tech. **14**, 140-147 (2003).

[27] D.M. King, N.J. Hangiandreou, D.J. Tradup, and S.F. Stekel. "Evaluation of a low-cost liquid ultrasound test object for detection of transducer artefacts." Phys Med Biol **55**, N557-570 (2010).

[28] J.M. Rubin, R.S. Adler, R.O. Bude, J.B. Fowlkes, and P.L. Carson. "Clean and dirty shadowing at US: a reappraisal." Radiol. **181**, 231-236 (1991).

[29] I. Koltsova, M. Deinega, and A. Polukhina. "Attenuation of ultrasound waves in suspensions of porous particles." Proc. of the Russian Acoust. Soc., 328-331 (2008).

[30] I. Koltsova, A. Polukhina, and M. Deynega. "Reversible and irreversible processes in biocomposites." Proc. of the Russian Acoust. Soc., 301-303 (2010).

[31] US National Research Council Int. Critical Tables of Numerical Data, Physics, Chemistry, and Technology III. 94 (1926-1933).

[32] E. White, M. Chellamuthu, and J. Rothstein. "Extensional rheology of a shear thickening cornstarch and water suspension." Rheol. Acta **49**, 119-129 (2009).

[33] I.Y. Kuo, B. Hete, and K.K. Shung. "A novel method for the measurement of acoustic speed." J. Acoust. Soc. Am. **88**, 1679-1682 (1990).

[34] B. Sollish. "A device for measuring ultrasonic propagation velocity in tissue." Nat. Bur. of Stand., Special Publication 525: Ultrason. Tissue Char. II. (1979).

[35] R. Bracewell. The Fourier Transform and its Applications, 3rd edition. (2000).

[36] W. Marczak, W. "Water as a standard in the measurements of speed of sound in liquids." J. Acoust. Soc. Am. **102**, 2776-2779 (1997).

[37] J. Ophir, T.H. Shawker, N.F. Maklad, J.G. Miller, S.W. Flax, P.A. Narayana, and J.P. Jones. "Attenuation estimation in reflection: progress and prospects." Ultrason. Imag. **6**, 349-395 (1984).

[38] W. Xu, and J. Kaufman. "Diffraction correction methods for insertion ultrasound attenuation estimates." IEEE Trans. on Biomed. Eng. **40**, 563-570 (1993).

[39] R.A. Urick. "Sound velocity method for determining the compressibility of finely divided substances." J. Appl. Phys. **18**, 983-987 (1947).

[40] K.K. Shung, B.A. Krisko, and J.O. Ballard. "Acoustic measurement of erythrocyte compressibility." J. Acoust. Soc. Am. **72**, 1364-1367 (1982).

[41] A. Wood. A Textbook of Sound. (1930).

[42] J. Allegra, and S. Hawley. "Attenuation of sound in suspensions and emulsions: theory and experiments." J. Acoust. Soc. Am. **51**, 1545-1564 (1972).

TABLE I. The speed of sound and attenuation coefficient at specific frequencies is displayed for each concentration of cornstarch in a 51.5% cesium chloride suspension.

| Cornstarch Mass Fraction | Speed of Sound (m/s) | Attenuation Coefficient (dB/cm) | | | | |
|---|---|---|---|---|---|---|
| | | 4 MHz | 5 MHz | 6 MHz | 7 MHz | 8 MHz |
| 0% | 1483 ± 10 | - | - | - | - | - |
| 10% | 1531 ± 6 | 1.4 ± 0.1 | 2.0 ± 0.1 | 2.6 ± 0.1 | 3.4 ± 0.1 | 4.2 ± 0.2 |
| 20% | 1602 ± 6 | 3.1 ± 0.3 | 4.3 ± 0.2 | 5.6 ± 0.2 | 7.0 ± 0.2 | 8.6 ± 0.2 |
| 30% | 1671 ± 6 | 4.9 ± 0.5 | 6.7 ± 0.5 | 8.6 ± 0.6 | 10.8 ± 0.6 | 13.2 ± 0.7 |
| 40% | 1765 ± 9 | 9.3 ± 1.1 | 12.0 ±1.2 | 15.0 ± 1.2 | 18.6 ± 1.3 | 22.7 ± 1.3 |

TABLE II. The speed of sound in pure cornstarch granules inferred from the speed of sound of suspensions. The bulk modulus was inferred from the determined speed of sound using Equation 6.

| Starch Suspension Mass and Volume Fraction | Measured Suspension Speed of Sound (m/s) | Inferred Starch Speed of Sound (m/s) | Inferred Starch Bulk Modulus (GPa) |
|---|---|---|---|
| 20% | 1602 ± 6 | 2800 ± 170 | 12 ± 1 |
| 30% | 1671 ± 6 | 2750 ± 90 | 12 ± 1 |
| 40% | 1765 ± 9 | 2880 ± 100 | 13 ± 1 |

**List of Figures**

FIG. 1       a) The experimental setup for the measurement of the speed of sound.  b) The experimental setup for the measurement of the attenuation coefficient as a function of frequency.

FIG. 2       The speed of sound of the cornstarch in a density-matched (51.5%) cesium chloride suspension.  The error bars are too small to be visible on this scale.

FIG. 3       The attenuation coefficient shown as a function of frequency a) on a linear scale and b) on a log-log scale.  Measurements were made of 8 independent samples of each concentration of cornstarch suspended in a 51.5% cesium chloride solution.

**Figure 1:**

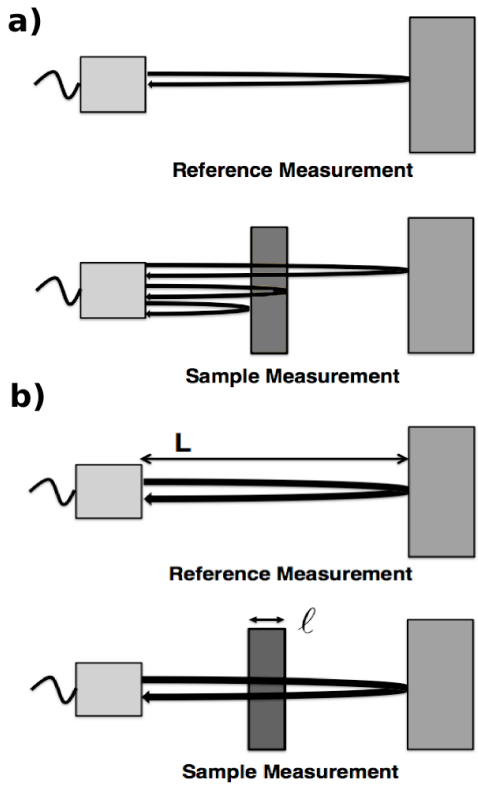

**Figure 2:**

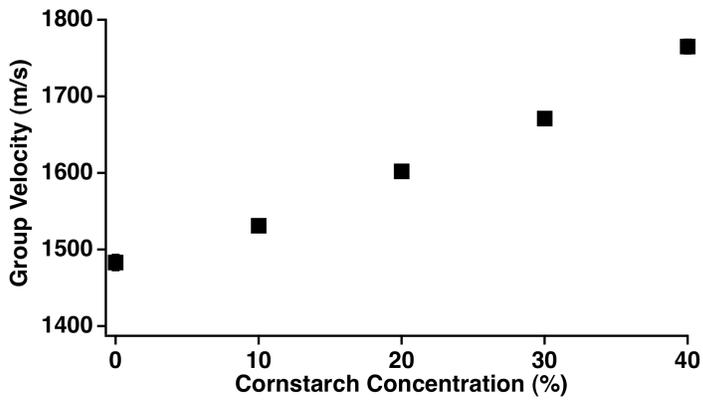

**Figure 3:**

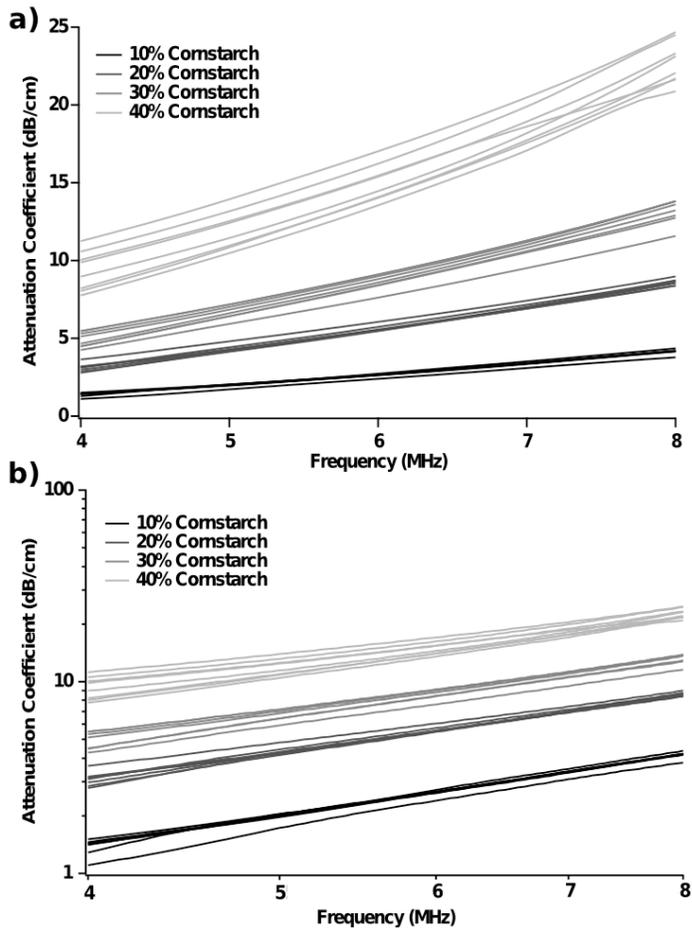